\documentclass[]{ws-ijmpb}
\usepackage{amsmath,amstext,amssymb,graphics,setspace,bm,dcolumn,color,psfrag}
\begin{document}

\markboth{Indrani Bose and Amit Kumar Pal}
{Quantum discord, decoherence and quantum phase transitions}

%
\catchline{}{}{}{}{}
%

\title{QUANTUM DISCORD, DECOHERENCE AND QUANTUM PHASE TRANSITION}

\author{INDRANI BOSE}

\address{Department of Physics, Bose Institute\\
93/1, Acharya Prafulla Chandra Road, Kolkata - 700009, India\\
indrani@bosemain.boseinst.ac.in}

\author{AMIT KUMAR PAL}

\address{Department of Physics, Bose Institute\\
93/1, Acharya Prafulla Chandra Road, Kolkata - 700009, India\\
ak.pal@bosemain.boseinst.ac.in}

\maketitle

\begin{abstract}
Quantum discord is a more general measure of quantum correlations than entanglement and has been proposed as a resource in certain quantum information
processing tasks. The computation of discord is mostly confined to two-qubit systems for which an analytical calculational scheme is available. The utilization of
quantum correlations in quantum information-based applications is limited by the
problem of decoherence, i.e., the loss of coherence due to the inevitable interaction
of a quantum system with its environment. The dynamics of quantum correlations
due to decoherence may be studied in the Kraus operator formalism for different
types of quantum channels representing system-environment interactions. In this
review, we describe the salient features of the dynamics of classical and quantum
correlations in a two-qubit system under Markovian (memoryless) time evolution.
The two-qubit state considered is described by the reduced density matrix obtained
from the ground state of a spin model. The models considered include the transverse-field $XY$ model in one dimension, a special case of which is the transverse-field 
Ising model, and the $XXZ$ spin chain. The
quantum channels studied include the amplitude damping, bit-flip, bit-phase-flip
and phase-flip channels. The Kraus operator formalism is briefly introduced and
the origins of different types of dynamics discussed. One can identify appropriate quantities associated with the dynamics of quantum correlations which provide
signatures of quantum phase transitions in the spin models. Experimental observations of the different types of dynamics are also mentioned.

\end{abstract}


\section{Introduction}
Interacting quantum systems are characterized by the presence of correlations between the different parts. 
The correlations are of two types: classical and quantum. Since the early days of quantum mechanics, the 
idea of entanglement has been invoked to probe the origins of quantum correlations \cite{epr}. In later years, 
various quantitative measures of entanglement have been proposed and the role of entanglement in a 
number of quantum information-based protocols highlighted \cite{vedralRMP,senAP,latorre}. A quantum state, pure or mixed, is either 
entangled or separable, the latter condition implying the absence of quantum correlations as measured by entanglement
in the state under consideration. In recent years, a new measure of quantum correlations, the quantum discord (QD), 
has been proposed based on the information theoretic concept of mutual information \cite{olivier,zurekRMP,henderson}. The QD is quantified by the 
difference between two quantum extensions of the classical mutual information. The two representations are 
identical in the classical domain.

In classical information theory, the total correlation between two random variables $A$ and $B$ is measured by their 
mutual information \cite{nielsen}
\begin{eqnarray}
 I(A,B)=H(A)+H(B)-H(A,B)
\label{mutual}
\end{eqnarray}
The random variables $A$ and $B$ take on the values \textquoteleft$a$\textquoteright$\;$ and \textquoteleft$b$\textquoteright$\;$
respectively with probabilities given by the sets $\left\{p_{a}\right\}$ and  $\left\{p_{b}\right\}$. The 
probability distribution $\left\{p_{ab}\right\}$ defines the outcome when joint measurements are carried out. The variables $A$  
and $B$ are correlated when $\left\{p_{ab}\right\}$
does not have a product form $\left\{p_{a}\times p_{b}\right\}$. In    
Eq. (\ref{mutual}), $H(A)=-\sum_{a}p_{a}\log_{2}p_{a}$,    $H(B)=-\sum_{b}p_{b}\log_{2}p_{b}$ and $H(A,B)=-\sum_{a,b}p_{ab}\log_{2}p_{ab}$
are the Shannon entropies for the variables $A$, $B$ and the joint system $AB$ respectively.  The probabilities $p_{a}$, $p_{b}$ and 
$p_{ab}$ satisfy the relations $p_{a}=\sum_{b}p_{ab}$ and $p_{b}=\sum_{a}p_{ab}$. The Shannon entropy of a random variable $A$ 
quantifies our ignorance about $A$ before we measure its value or equivalently it provides a measure of how much information is gained on an
average after a measurement is carried out \cite{nielsen}. An alternative representation of the classical mutual information is given by \cite{sarandy,luo}
\begin{eqnarray}
 J(A,B)=H(A)-H(A|B)
\label{mutual2}
\end{eqnarray}
where $H(A|B)$ is the conditional entropy and quantifies our lack of knowledge of the value of $A$ when that of $B$ is known. The exact 
equivalence of the expressions in Eqs. (\ref{mutual}) and (\ref{mutual2}) can be demonstrated using the Bayes rule $p_{ab}=p_{a|b}\,p_{b}$ 
and the definition $H(A|B)=-\sum_{a,b}p_{ab}\log_{2}p_{a|b}$ of the conditional entropy.

The generalization of the classical mutual information to the quantum case is achieved by replacing the classical probability distribution
and the Shannon entropy by the density matrix $\rho$ and the von Neumann entropy $S(\rho)=-\mbox{tr}\left(\rho\log_{2}\rho\right)$, 
respectively. The quantum generalizations of Eqs. (\ref{mutual}) and (\ref{mutual2}) are given by 
\begin{eqnarray}
 I\left(\rho_{AB}\right)&=&S\left(\rho_{A}\right)+S\left(\rho_{B}\right)-S\left(\rho_{AB}\right)\label{qmu}\\
 J\left(\rho_{AB}\right)&=&S\left(\rho_{A}\right)-S\left(\rho_{A}|\rho_{B}\right)\label{qmu2} 
\end{eqnarray}
where $S\left(\rho_{AB}\right)$ is the quantum joint entropy and $S\left(\rho_{A}|\rho_{B}\right)$ the quantum conditional entropy. The 
latter quantity is, however, ambiguously defined as the magnitude of the quantum conditional entropy (ignorance of $A$ once $B$ is known)
depends explicitly on the type of measurement carried out on $B$. Since different measurement choices yield different results, Eqs. (\ref{qmu})
and (\ref{qmu2})  are no longer identical. We consider von Neumann-type measurements on $B$ defined in terms of a complete set of orthogonal 
projectors $\left\{\Pi_{i}^{B}\right\}$ corresponding to the set of possible outcomes $i$. The state of the system after the measurement is given by 
\begin{eqnarray}
 \rho_{i}=\left(I\otimes\Pi_{i}^{B}\right)\rho_{AB}\left(I\otimes\Pi_{i}^{B}\right)/p_{i}
\end{eqnarray}
with 
\begin{eqnarray}
 p_{i}=tr \left(\left(I\otimes\Pi_{i}^{B}\right)\rho_{AB}\left(I\otimes\Pi_{i}^{B}\right)\right)
\end{eqnarray}
$I$ denotes the identity operator for the subsystem $A$ and $p_{i}$
is the probability of obtaining the outcome $i$. From Eq. (\ref{qmu2}), an alternative expression of quantum mutual information is given by \cite{sarandy,luo}
\begin{eqnarray}
 J\left(\rho_{AB},\left\{ \Pi_{i}^{B}\right\} \right)
=S\left(\rho_{A}\right)-S\left(\rho_{AB}\left|\left\{ \Pi_{i}^{B}\right\} \right.\right)
\label{qmu3}
\end{eqnarray}
The quantum analogue of the conditional entropy is 
\begin{equation}
S\left(\rho_{AB}\left|\left\{ \Pi_{i}^{B}\right\} \right.\right)
=\sum_{i}p_{i}S\left(\rho_{i}\right)
\end{equation}
Henderson and Vedral \cite{henderson} have shown that the maximum of $J\left(\rho_{AB},\left\{ \Pi_{i}^{B}\right\} \right)$  w.r.t. $\left\{\Pi_{i}^{B}\right\}$ 
provides a measure of the classical correlations (CC), $C\left(\rho_{AB}\right)$, i.e., 
\begin{equation}
C\left(\rho_{AB}\right)
=\underset{\left\{ \Pi_{i}^{B}\right\} }{\mbox{max}}
\left(J\left(\rho_{AB},\left\{ \Pi_{i}^{B}\right\} \right)\right)
\label{class}
\end{equation}
The difference between the total correlations 
$I\left(\rho_{AB}\right)$ (Eq. (\ref{qmu})) and the CC,  $C\left(\rho_{AB}\right)$, defines the QD, $Q\left(\rho_{AB}\right)$. 
\begin{eqnarray}
 Q\left(\rho_{AB}\right)=I\left(\rho_{AB}\right)-C\left(\rho_{AB}\right)
\label{qd}
\end{eqnarray}
The QD is defined for bipartite systems only and due to the computational difficulty of carrying out the extremization process in Eq. (\ref{class}), the 
calculation of the QD is mostly confined to two-qubit systems. For any pure state, the QD reduces to the entropy of entanglement \cite{luo} and the 
total correlations, measured by the mutual information, are equally divided between the classical and quantum correlations, i.e., 
$C\left(\rho_{AB}\right)=Q\left(\rho_{AB}\right)=\frac{1}{2}I\left(\rho_{AB}\right) $. 
      
In the case of mixed states, however, the QD and the entanglement provide different measures of quantum correlations. In fact, there are mixed states 
which are separable, i.e., unentangled but  for which QD is non-zero. There is thus a shift in focus in the case of the QD from the separability versus 
entanglement criteria to the issue of classical versus quantum correlations. The relationship between the QD, entanglement and classical 
correlations is not as yet clearly understood even for the simple two-qubit system. The QD is not always larger than entanglement and is not always 
less than classical correlations \cite{ali}. The QD defined by Eqs. (\ref{qmu}), (\ref{class}) and (\ref{qd}) does not provide a measure of quantum correlations 
in multipartite systems. Recently, the concept of relative entropy has been utilized to obtain measures of classical and non-classical correlations in a 
given quantum state \cite{modi,modi2}. The relative entropy between two quantum states $x$ and $y$ is given by $S(x||y)=\mbox{tr}\left(x\log_{2}\frac{x}{y}\right)$. 
It is a non-negative quantity and hence serves as a \textquotedblleft distance\textquotedblright$\;$ measure of the state $x$ from the state $y$. The relative 
entropy of entanglement is the distance of the entangled state $\rho$ to the closest separable state $\sigma$. Similarly, the discord is measured by the 
distance between the quantum state and its closest classical state. Analogous definitions exist for quantum dissonance (non-classical correlations for 
separable states), total mutual information and classical correlations. The measures are valid in the multipartite case and for arbitrary dimensions 
of the system. Since the different types of correlations have a common measure, additivity relations connecting the correlations can be derived. Other 
alternative measures of non-classical correlations include the geometric QD \cite{dakic} and the Gaussian QD \cite{giorda}. In this review, we confine ourselves to 
the original definition of the QD as embodied in Eqs. (\ref{qmu}), (\ref{class}) and (\ref{qd}). 

Condensed matter systems like molecular magnets (represented by spin clusters) and spin chains have been extensively studied to characterize as well as quantify 
the quantum correlations present in the ground and thermal states of the $\,$ spin systems \cite{vedralRMP,senAP,modi3,celeri}. While the bulk of the studies are devoted to entanglement in its 
various forms, there are now several studies on the QD properties of well-known $\,$  spin models
\cite{henderson,modi3,celeri,dillenschneider,pal1,werlang,maziero2,werlang2,maziero3,libherti,hassan,tomasello,dhar,pal2,chen,tian,yurischev}. Many of these studies focus on how the QD and 
associated quantities provide signatures of quantum phase transitions in specific spin systems. Quantum phase transitions (QPTs) in interacting systems occur at 
$T=0$ and are brought about by tuning  a non-thermal parameter $g$, e.g., pressure, chemical composition or external magnetic field, to a special value $g_{c}$
\cite{sachdev,sachdev2}. The transition is driven by quantum fluctuations and brings about qualitative changes in the ground state wave function at the transition point. It is then 
reasonable to expect that the quantum correlations present in the ground state would provide signatures of the occurrence of a QPT. Apart from the effect of 
changing parameters on the quantum correlations of a system, the inevitable interaction of the system with its environment results in decoherence, i.e., a 
destruction of quantum properties including correlations \cite{nielsen}. The dynamics of entanglement and QD under system-environment interactions have been investigated 
in a number of $\,$ recent studies \cite{pal1,pal2,maziero4,almeida,werlang3,maziero5,mazzola1,pal3,ferraro}. One feature which emerges out of such studies is that the QD is more robust than entanglement in the case of 
Markovian (memoryless) time evolution. The dynamics may bring about the sudden disappearance of entanglement at a finite time termed the \textquoteleft entanglement
sudden death\textquoteright$\;$ \cite{maziero4,almeida}. The QD, on the other hand, decays in time but vanishes only asymptotically \cite{pal1,pal2,werlang3,mazzola1,pal3,ferraro}. 
Also, under Markovian time evolution and for 
a class of initial states, the decay rates of the classical and quantum correlations exhibit sudden changes \cite{maziero5,mazzola1}. In two recent studies, the dynamics of the mutual 
information $I\left(\rho_{AB}\right)$, the classical correlations $C\left(\rho_{AB}\right)$ and the quantum correlations $Q\left(\rho_{AB}\right)$, as measured 
by the QD, have been studied in two-qubit states the density matrices of which are the reduced density matrices obtained from the ground states of the transverse-field 
Ising model (TIM) and the transverse-field $XY$ model in one dimension (1d) \cite{pal2,pal3}. The time evolution brought about by system-environment interactions is assumed to be Markovian in 
nature and the quantum channels, representing qubit-environment interactions, include amplitude damping (AD), bit-flip (BF), phase-flip (PF) and bit-phase-flip (BPF). A significant outcome 
of the studies is the identification of appropriate quantities associated with the dynamics of the correlations which signal the occurrence of a QPT. The TIM in 1d is a 
special case of the transverse-field $XY$ model in 1d. The latter model has a rich phase diagram exhibiting QPTs \cite{mattis,mccoy,pfeuty,zhong,dutta,its}. 
In fact, both the spin models are well-known 
statistical mechanical models which illustrate QPTs. In this review, we combine the themes of quantum correlations, decoherence and QPTs in the study of spin 
models like the TIM in 1d, the transverse-field $XY$ chain and the $XXZ$ spin chain. In Section 2, we describe the spin models and the QPTs associated with them. 
In Section 3, we introduce the Kraus operator formalism for describing the time evolution of open quantum systems, i.e., systems interacting with specific environments. Some 
representative quantum channels are described as also the dynamics of mutual information, classical and quantum correlations. Section 4 describes the main results obtained 
so far \cite{pal2,pal3} as well as some new results. The methodology reported in the review is general in nature and applicable to other condensed matter models exhibiting QPTs. In Section 5, some 
concluding remarks are made and future research directions pointed out. 

\section{Quantum Phase Transitions in Spin Models}

\begin{figure}
\begin{center}
 \includegraphics[scale=0.45]{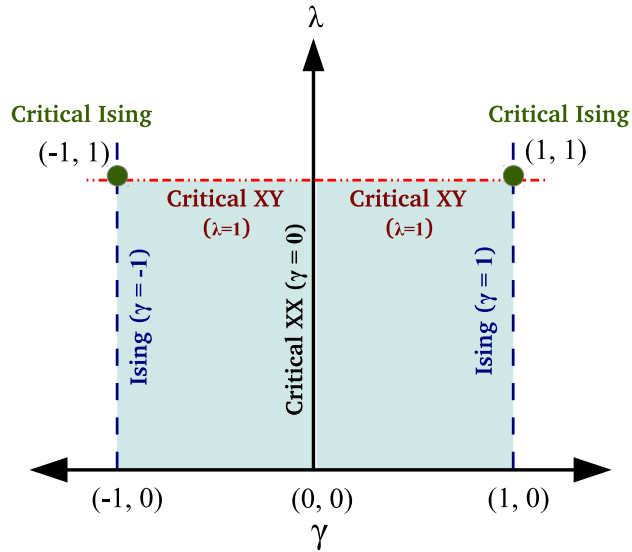}
\end{center}
\caption{Criticality of the transverse-field $XY$ model in the $(\gamma,\lambda)$ parameter space. 
         The transverse-field Ising model
         $(\gamma=\pm1)$ has a QCP at $\lambda_{c}=1$. The line $\lambda=1$ 
         represents criticality of the $XY$ model with universality class the same as that of 
         the transverse-field Ising model.  
         The $\gamma=0$ line ($XX$ model) represents the anisotropy transition line
         for $\lambda\in[0,1]$ with the transition belonging to a new universality class.}
\label{phd}
\end{figure}
The fully anisotropic Heisenberg spin chain in a magnetic field is described by the Hamiltonian 
\begin{eqnarray}
 H_{XYZ}=-\sum_{i=1}^{L}\left[J_{x}\sigma_{i}^{x}\sigma_{i+1}^{x}+J_{y}\sigma_{i}^{y}\sigma_{i+1}^{y}+J_{z}\sigma_{i}^{z}\sigma_{i+1}^{z}\right]-h\sum_{i=1}^{L}\sigma_{i}^{z}
\label{hxxz}
\end{eqnarray}
where $\sigma_{i}^{\alpha}\;\left(\alpha=x,y,z\right)$ is the Pauli spin operator at the $i$th site, $J_{x}$, $J_{y}$ and $J_{z}$ are the strengths of the nearest-neighbour 
(n.n.) exchange interactions and $h$ the external magnetic field. $L$ denotes the total number of spins in the chain. A number of spin models are special cases of the fully 
anisotropic model: \textit{(i)} $XXZ$ spin model in a magnetic field $\left(J_{x}=J_{y}\neq J_{z},\;h\neq 0\right)$, \textit{(ii)} transverse-field $XY$ model    
$\left(J_{x}\neq J_{y},\;J_{z}=0,\;h\neq 0\right)$, \textit{(iii)} transverse-field $XX$ model $\left(J_{x}= J_{y},\;J_{z}=0,\;h\neq 0\right)$ and \textit{(iv)} the TIM
$\left(J_{y}=J_{z}=0,\;h\neq 0\right)$. One can also consider cases for which $h=0$. The TIM in this case reduces to the Ising model which has no quantum character. 

The transverse-field $XY$ model in 1d describes an interacting spin system for which many exact results on the ground and excited state properties including spin correlations 
are known \cite{mattis,mccoy,pfeuty}. The corresponding Hamiltonian is written as 
\begin{eqnarray}
H_{XY}=-\frac{\lambda}{2}\sum_{i=1}^{L}
\left\{(1+\gamma)\sigma_{i}^{x}\sigma_{i+1}^{x}+(1-\gamma)\sigma_{i}^{y}\sigma_{i+1}^{y}\right\}
-\sum_{i=1}^{L}\sigma_{i}^{z}
\label{hxy}
\end{eqnarray}
where $\gamma$ is the degree of anisotropy ($-1\leq\gamma\leq1$) and $\lambda$ is inversely proportional to the strength of the transverse magnetic field in the $z$ direction $(\lambda>0)$. The 
Hamiltonian satisfies periodic boundary condition, i.e., $L+1\equiv 1$ and is translationally invariant. Two special cases of the $XY$ model are the TIM  with $\gamma=\pm1$ and the 
isotropic $XX$ model $(\gamma=0)$ in a transverse magnetic field. For the full range of values of the anisotropic parameter, $H_{XY}$ can be diagonalized exactly in the 
thermodynamic limit $L\rightarrow\infty$ \cite{mattis,mccoy,pfeuty}. This is achieved via the successive applications of the Jordan-Wigner and Bogoliubov transformations. Figure \ref{phd} 
describes the critical transitions associated with the transverse-field $XY$ model in the $(\gamma,\lambda)$ parameter space. For non-zero values of $\gamma$, a second-order QPT
occurs at the critical point $\lambda_{c}=1$ separating a ferromagnetic ordered phase ($\lambda>1$) from a quantum paramagnetic phase ($\lambda<1$). The transition is characterized by the order parameter 
$\left\langle\sigma^{x}\right\rangle$, the magnetization in the $x$ direction, which has a non-zero expectation value only in the ordered ferromagnetic phase $(\lambda>1)$. 
The magnetization in the $z$ direction, $\left\langle\sigma^{z}\right\rangle$, is non-zero for all values of $\lambda$ with its first derivative exhibiting a singularity at 
the critical point  $\lambda_{c}=1$. When $0<|\gamma|\leq1$, the critical point transition occurring at $\lambda_{c}=1$ belongs to the Ising universality class. For $\lambda\in[0,1]$,
there is another QPT, termed the anisotropy transition, at the critical point $\gamma=0$. The transition belongs to a different universality class and separates two ferromagnetic phases 
with orderings in the $x$ and $y$ directions respectively \cite{mattis,mccoy,pfeuty,zhong,dutta}. The TIM Hamiltonian in 1d is obtained from Eq. (\ref{hxy}) by putting $\gamma=1$ and is given by 
\begin{eqnarray}
 H_{TIM}=-\lambda\sum_{i=1}^{L}\sigma_{i}^{x}\sigma_{i+1}^{x}-\sum_{i=1}^{L}\sigma_{i}^{z}
\label{htim}
\end{eqnarray}
When the parameter $\lambda=0$, all the spins are oriented in the positive $z$ direction in the ground state whereas in the extreme limit $\lambda=\infty$ the ground state is doubly 
degenerate with all the spins pointing in either the positive or the negative $x$ direction. As mentioned before, a QPT occurs at the critical point $\lambda_{c}=1$ with 
$\left\langle\sigma^{x}\right\rangle\neq0$ in the ordered ferromagnetic phase $(\lambda>1)$. The XXZ spin chain in a zero magnetic field has the Hamiltonian 
\begin{eqnarray}
 H_{XXZ}=-\frac{J}{2}\sum_{i=1}^{L}\left[\sigma_{i}^{x}\sigma_{i+1}^{x}+\sigma_{i}^{y}\sigma_{i+1}^{y}+\Delta\sigma_{i}^{z}\sigma_{i+1}^{z}\right]
\label{hxxz2}
\end{eqnarray}
Two QPTs, one first-order and the other infinite-order, occur at the respective points $\Delta=1$ (ferromagnetic point) and $\Delta=-1$ (antiferromagnetic point). At the latter point, the $XXZ$ spin chain 
undergoes a QPT from an antiferromagnetic phase $(\Delta<-1)$ to an $XY$ phase for $-1<\Delta<1$. 

In recent years, quantum information-related measures like entanglement, discord and fidelity \cite{vedralRMP,senAP,sarandy,dillenschneider,osborne,osterloh,bose,zanardi,tribedi} 
have been shown to provide signatures of QPTs. An $n$th order QPT is characterized by a 
discontinuity/divergence in the $n$th derivative of the ground state energy with respect to the tuning parameter $g$ as $g\rightarrow g_{c}$, defining the transition point. At a first-order QPT
point, appropriate entanglement measures and the QD are known to become discontinuous \cite{sarandy,dillenschneider,dutta,maziero2} whereas a second-order (critical point) QPT is signaled by the 
discontinuity or divergence of the 
first derivative of the quantum correlation measures with respect to the tuning parameter at the critical point \cite{sarandy,dillenschneider,maziero2}. At a quantum critical point, quantum fluctuations occur on all length scales leading 
to a divergent correlation 
length. The ground state energy and related quantities become non-analytic as the tuning parameter $g$ tends to the critical point $g_{c}$. The influence of a QPT extends into the finite temperature 
part of the phase diagram so that experimental detection of the QPT is possible. 

We now quote some major results on the signatures of QPTs provided by quantum correlation measures like entanglement and the QD in the cases of spin models like 
the TIM in 1d, transverse-field $XY$ and $XXZ$ spin chains. For more detailed information, the reader is referred to the original papers and reviews \cite{vedralRMP,senAP,modi3,celeri,bose,dutta}. 
In the case of the TIM in 1d, the n.n. 
concurrence, a measure of pairwise entanglement, reaches its maximum value close to the QCP $\lambda_{c}=1$. The next-nearest-neighbour (n.n.n.) concurrence, on the other hand, has its maximum value at 
$\lambda_{c}=1$. The first derivative of the n.n. concurrence w.r.t. the tuning parameter $\lambda$ exhibits a logarithmic divergence as $\lambda\rightarrow\lambda_{c}$. The pairwise entanglement 
as measured by the concurrence does not become long-ranged as $\lambda\rightarrow\lambda_{c}$, in fact it falls to zero beyond the n.n.n. distance.  Both the n.n. and the n.n.n. QD attain their maximal
values close to the critical point $\lambda_{c}$. The first derivative of the n.n. (n.n.n.) QD with respect to $\lambda$ becomes discontinuous (singular) at the critical point $\lambda_{c}=1$. The magnitude 
of the classical correlations has a monotonic dependence on $\lambda$ starting with low values for small $\lambda$. The transverse-field $XY$ chain belongs to the universality class of the TIM in 1d away 
from the isotropic limit $\gamma=0$. In the case of the zero-field $XXZ$ chain, the n.n. concurrence is maximal at the infinite-order critical point $\Delta_{c}=-1$. The n.n. QD is found to be maximal 
(with a discontinuity) and the classical correlations become minimal (with a kink) at $\Delta_{c}=-1$. At the first-order transition point $\Delta=1$, the n.n. classical and quantum correlations are 
discontinuous \cite{sarandy,dillenschneider}. For the transverse-field $XY$ chain, Maziero et. al. \cite{maziero2} derived analytic expressions for classical and quantum correlations for spin-pairs separated by an arbitrary 
distance for both temperatures $T=0$ and $T\neq 0$. The $T=0$ QD for spin pairs beyond the n.n.n. distance is able to signal a QPT whereas pairwise entanglement for the same spin pairs is unable to do so. 
Pairwise entanglement is typically short-ranged even close to criticality. Spin chains  like the transverse-field $XY$ chain and the $XXZ$ chain in the presence of domain walls are characterized by a long-range 
decay of the QD as a function of the spin-spin distance close to the quantum critical point \cite{maziero3}. The decay rate of the QD has a noticeable change as the critical point is crossed. In a recent study 
\cite{werlang2}, the thermal QD is shown to be maximal and its first derivative with respect to the tuning parameter discontinuous at the quantum critical point $\Delta=-1$ of the $XXZ$ chain for both 
$T=0$ and $T>0$. The entanglement of formation is maximal at the critical point only for $T=0$, the maximum shifts from $\Delta=-1$ for $T>0$. The thermal QD also signals the QPT at $\Delta=1$. Thus, the QD 
has the important property of being able to detect the critical points of QPTs at finite temperatures. In contrast, both entanglement and thermodynamic quantities fail to signal QPTs for $T>0$. The special 
property of the QD could be useful in the experimental detection of critical points. 

\section{Dynamics of Correlations}         
We next consider the interaction of the chain of qubits (spins) with 
an environment. The initial state of the whole system at time $t=0$ 
is assumed to be of the product form, i.e., 
\begin{eqnarray}
 \rho(0)=\rho_{s}(0)\otimes\rho_{e}(0)
 \label{sysden}
\end{eqnarray}
with the density matrices $\rho_{s}$ and $\rho_{e}$ corresponding to the system
(spin chain) and the environment 
respectively. We assume that the environment is described in terms of $L$ independent 
reservoirs each of which interacts locally with a qubit constituting the spin chain. 
The two-qubit reduced density matrices, $\rho_{rs}$ and $\rho_{re}$,
are obtained by taking
partial traces on $\rho_{s}$ and $\rho_{e}$ respectively over the states of all the qubits other than the two chosen qubits. The two-qubit reduced density matrix $\rho_{r}(0)$ obtained from Eq. (\ref{sysden})
can be written as 
 \begin{eqnarray}
 \rho_{r}(0)=\rho_{rs}(0)\otimes\rho_{re}(0)
\label{redden}
\end{eqnarray}
The quantum channel describing the interaction between a qubit and its local environment can be of various types: AD, phase damping, BF, PF, BPF etc. \cite{nielsen,zanardi}.
With the initial reduced density matrix given in Eq. (\ref{redden}), the objective is to determine the dynamics of the two-qubit classical and quantum correlations (in the form of the QD) under the influence of various
quantum channels. The time evolution of the closed quantum system consisting of both the system and the environment is given by 
 \begin{eqnarray}
 \rho_{se}(t)=U(t)\rho_{se}(0)U^{\dagger}(t)
 \label{evolve2}
\end{eqnarray}
where $U(t)$ is the unitary evolution operator generated by the total Hamiltonian $H$
$\left(U=e^{-iHt/\hbar}\right)$. $H$ is written as $H=H_{s}+H_{e}+H_{se}$ where $H_{s}$ 
and $H_{e}$ are the bare system and the environment Hamiltonians respectively and $H_{se}$ 
the Hamiltonian describing the interactions between the system and the environment. The time 
evolution of the system $s$ under the influence of the environment $e$ is obtained by 
carrying out a partial trace on $\rho_{se}(t)$ (Eq. (\ref{evolve2})) over the environment 
states, i.e.,
\begin{eqnarray}
 \rho_{s}(t)=\mbox{Tr}_{e}\left[U(t)\rho_{se}(0)U^{\dagger}(t)\right]
\label{partr}
\end{eqnarray}
In Eq. (\ref{partr}), $\rho_{se}(0)=\rho(0)$ from Eq. (\ref{sysden}). Let $|e_{k}\rangle$ be an orthogonal basis spanning the finite-dimensional state space of the 
environment. With the initial state of the whole system given by Eq. (\ref{sysden}),
\begin{eqnarray}
 \rho_{s}(t)=\sum_{k}\langle e_{k}|U\left[\rho_{s}(0)\otimes\rho_{e}(0)\right]U^{\dagger}|e_{k}\rangle
\end{eqnarray}

\begin{figure}
\begin{center}
 \includegraphics[scale=0.6]{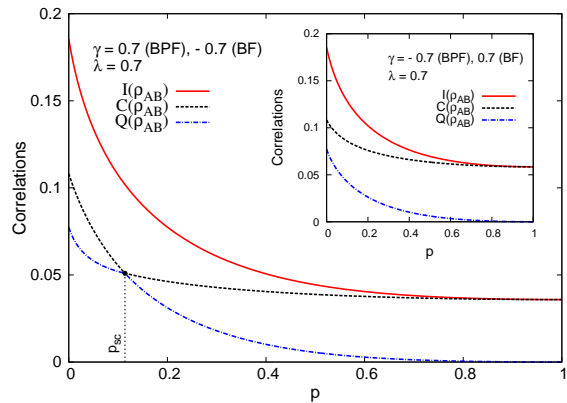}
\end{center}
\caption{BPF and BF channel: Decay of mutual information 
         $I\left(\rho_{AB}\right)$ (solid line), classical correlations $C\left(\rho_{AB}\right)$
         (dashed line), and the QD $Q\left(\rho_{AB}\right)$ (dot-dashed line) as a function of the parametrized time 
         $p=1-e^{-\theta t}$ for $\lambda=0.7,\gamma=0.7$ (BPF channel) and $\lambda=0.7,\gamma=-0.7$ (BF channel). 
         Also, $p_{sc}=0.114$. (inset) Decay of mutual information $I\left(\rho_{AB}\right)$ (solid line), classical correlations $C\left(\rho_{AB}\right)$ 
         (dashed line) and quantum discord $Q\left(\rho_{AB}\right)$ (dot-dashed line) as a function of the parametrized time $p=1-e^{-\theta t}$ for $\lambda=0.7,\gamma=-0.7$ (BPF channel)
         and $\lambda=0.7,\gamma=0.7$ (BF channel).}
\label{dyn}
\end{figure}

Let $\rho_{e}(0)=|e_{0}\rangle\langle e_{0}|$ be the initial state of the environment. Then
\begin{eqnarray}
 \rho_{s}(t)=\sum_{k}E_{k}\rho_{s}(0) E_{k}^{\dagger}
 \label{sysev}
\end{eqnarray}
where $E_{k}\equiv\langle e_{k}|U|e_{0}\rangle$ is the Kraus operator which acts on the state space of 
the system only \cite{nielsen,maziero4}. Let $\left\{\phi_{i}\right\}$, $i=1,2,...,d,$ define 
the basis in the state space of the system $s$. There are then at most $d^{2}$ independent 
Kraus operators $E_{k}$, $k=0,...,d^{2}-1$ \cite{nielsen,salles}. The unitary evolution of 
$s+e$ is given by the map:
\begin{eqnarray}
 |\phi_{1}\rangle|e_{0}\rangle &\rightarrow& E_{0}|\phi_{1}\rangle|e_{0}\rangle+...
                                             +E_{d^{2}-1}|\phi_{1}\rangle|e_{d^{2}-1}\rangle \nonumber \\
 |\phi_{2}\rangle|e_{0}\rangle &\rightarrow& E_{0}|\phi_{2}\rangle|e_{0}\rangle+...
                                             +E_{d^{2}-1}|\phi_{2}\rangle|e_{d^{2}-1}\rangle \nonumber \\
 &\vdots& \nonumber \\
 |\phi_{d}\rangle|e_{0}\rangle &\rightarrow& E_{0}|\phi_{d}\rangle|e_{0}\rangle+...
                                             +E_{d^{2}-1}|\phi_{d}\rangle|e_{d^{2}-1}\rangle 
\label{map1}
\end{eqnarray}
In compact notation, the map is given by 
\begin{eqnarray}
 U|\phi_{i}\rangle|e_{0}\rangle\equiv\sum_{k}E_{k}|\phi_{i}\rangle|e_{k}\rangle, \; i=1,2,...,d
\label{map2}
\end{eqnarray}
In the case of $N$ system parts with each part interacting with a local independent environment, 
Eq. (\ref{sysev}) becomes
\begin{eqnarray}
 \rho_{s}(t)=\sum_{k_{1},..,k_{N}}E_{k_{1}}^{(1)}\otimes..\otimes E_{k_{N}}^{(N)}\rho_{s}(0)
             E_{k_{1}}^{(1)\dagger}\otimes..\otimes E_{k_{N}}^{(N)\dagger}\nonumber \\
 \;\;\;
\label{npart}
\end{eqnarray}
where $E_{k_{\alpha}}^{(\alpha)}$ is the $k_{\alpha}$th Kraus operator with the environment acting on the system part 
$\alpha$. The specific form for $\rho_{s}(t)$ arises as the total evolution operator can be written 
as $U(t)=U_{1}(t)\otimes U_{2}(t) \otimes ... \otimes U_{N}(t)$.
Following the general formalism of the Kraus operator representation, an initial state, 
$\rho_{rs}(0)$, of the two-qubit reduced density matrix evolves as \cite{nielsen,werlang3}
\begin{eqnarray}
 \rho_{rs}(t)=\sum_{\mu,\nu}E_{\mu,\nu}\rho_{rs}(0)E_{\mu,\nu}^{\dagger}
\label{evolve}
\end{eqnarray}
where the Kraus operators $E_{\mu,\nu}=E_{\mu}\otimes E_{\nu}$ satisfy the completeness relation
$\sum_{\mu,\nu}E_{\mu,\nu}E_{\mu,\nu}^{\dagger}=I$ for all $t$. We now briefly describe the various 
quantum channels and write down the corresponding Kraus operators. A fuller 
description can be obtained from Refs. [\refcite{nielsen,werlang3}]. 

\noindent (i) \textit{AD Channel.}
The channel describes the dissipative interaction between a system and its environment resulting in 
an exchange of energy between $s$ and $e$ so that $s$ is ultimately in thermal equilibrium with 
$e$. The  $s+e$ time evolution is given by the unitary transformation
\begin{eqnarray}
 |0\rangle_{s}|0\rangle_{e} &\rightarrow& |0\rangle_{s}|0\rangle_{e}
\label{map31}
\end{eqnarray}
\begin{eqnarray}
 |1\rangle_{s}|0\rangle_{e} &\rightarrow& \sqrt{q}|1\rangle_{s}|0\rangle_{e}+\sqrt{p}|0\rangle_{s}|1\rangle_{e}
 \label{map32}
\end{eqnarray}
where $|0\rangle_{s}$ and $|1\rangle_{s}$ are the ground and excited qubit states and $|0\rangle_{e}$,
$|1\rangle_{e}$ denote states of the environment with no excitation (vacuum state) and one excitation
respectively. Eq. (\ref{map31}) stipulates that there is no dynamic evolution if the system 
and the environment are in their ground states. Eq. (\ref{map32}) states that if the system qubit 
is in the excited state, the probability to remain in the same state is $q$ and the probability 
of decaying to the ground state is $p$ $(p+q=1)$. The decay of the qubit state is accompanied by 
a transition of the environment to a state with one excitation. The qubit states may be two atomic 
states with the excited state decaying to the ground state by emitting a photon. The environment
on acquiring the photon is no longer in the vacuum state. With a knowledge of the map equations 
(Eqs. (\ref{map31}) and (\ref{map32})), the Kraus operators for the AD channel 
can be written as
\begin{eqnarray}
 E_{0}=\left(
 \begin{array}{cc}
  1 & 0 \\
  0 & \sqrt{q}
 \end{array}\right);\;\;
 E_{1}=\left(
 \begin{array}{cc}
  0 & \sqrt{p} \\
  0 & 0
 \end{array}\right)
\end{eqnarray}
where $q=1-p$. The Kraus operators for the two distinct environments (one for each qubit) have 
identical forms. In the case of Markovian time evolution, $p$ is given by  $p=1-e^{-\theta t}$ 
with $\theta$ denoting the decay rate.    

\noindent (ii) \textit{Phase Damping (dephasing) Channel.}
The channel describes the loss of quantum coherence without loss of energy. The Kraus operators are:
\begin{eqnarray}
 E_{0}=\left(
 \begin{array}{cc}
  1 & 0 \\
  0 & \sqrt{q}
 \end{array}\right);\;\;
 E_{1}=\left(
 \begin{array}{cc}
  0 & 0 \\
  0 & \sqrt{p}
 \end{array}\right)
\end{eqnarray}  
with $q=1-p$ and $p=1-e^{-\theta t}$.

\begin{figure}
\begin{center}
 \includegraphics[scale=0.6]{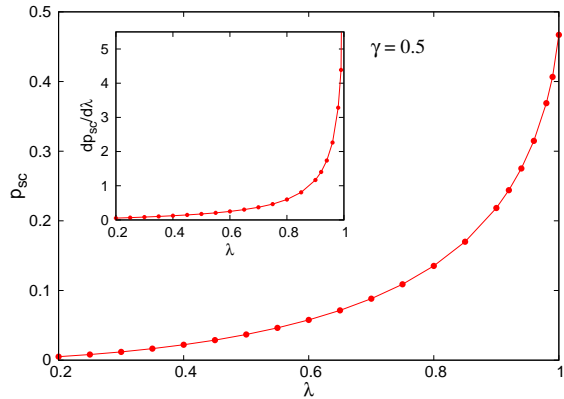}
\end{center}
\caption{Variation of $p_{sc}$ as a function of $\lambda$ with $\gamma=0.5$ for the BPF channel. (inset) The first derivative of $p_{sc}$ w.r.t. $\lambda$ diverges as 
         the QCP $\lambda_{c}=1$ is approached.}
\label{qpt}
\end{figure}

\noindent (iii) \textit{BF, PF and BPF channels.}      
The channels destroy the information contained in the phase relations without involving an exchange 
of energy. The Kraus operators are
\begin{eqnarray}
 E_{0}=\sqrt{q^{\prime}}\left(
 \begin{array}{cc}
  1 & 0 \\
  0 & 1
 \end{array}\right);\;\;
 E_{1}=\sqrt{p/2}\sigma_{i}
\end{eqnarray}  
where $i=x$ for the BF, $i=y$ for the BPF and $i=z$ for the PF channel with
$q^{\prime}=1-p/2$ and  $p=1-e^{-\theta t}$. The expanded forms of the Kraus operators are:

\noindent BF:
\begin{eqnarray}
  E_{0}&=&\left(
 \begin{array}{cc}
  \sqrt{1-p/2} & 0 \\
  0 & \sqrt{1-p/2}
 \end{array}\right) \nonumber \\
 E_{1}&=&\left(
 \begin{array}{cc}
  0 & \sqrt{p/2} \\
  \sqrt{p/2} & 0
 \end{array}\right)
\label{kbf}
\end{eqnarray}
\noindent PF:
\begin{eqnarray}
  E_{0}&=&\left(
 \begin{array}{cc}
  \sqrt{1-p/2} & 0 \\
  0 & \sqrt{1-p/2}
 \end{array}\right)\nonumber \\
 E_{1}&=&\left(
 \begin{array}{cc}
  \sqrt{p/2} & 0 \\
  0 & -\sqrt{p/2}
 \end{array}\right)
\label{kpf}
\end{eqnarray}
\noindent BPF: 
\begin{eqnarray}
  E_{0}&=&\left(
 \begin{array}{cc}
  \sqrt{1-p/2} & 0 \\
  0 & \sqrt{1-p/2}
 \end{array}\right) \nonumber \\
 E_{1}&=&\left(
 \begin{array}{cc}
  0 & -i\sqrt{p/2} \\
  i\sqrt{p/2} & 0
 \end{array}\right)
\label{kbpf}
\end{eqnarray}
As shown in Ref. [\refcite{nielsen}], the phase damping quantum operation is identical to that of 
the PF channel so that we will consider only one of these, the PF channel, in the 
following. 

In the case of the transverse-field $XY$ chain, including the TIM, the reduced density matrix $\rho_{AB}$ has the form of $X$-states with 
\begin{eqnarray}
\rho_{AB}= \left(
 \begin{array}{cccc}
 a & 0 & 0 & f \\
 0 & b & z & 0 \\
 0 & z & b & 0 \\
 f & 0 & 0 & d 
 \end{array}
 \right)
 \label{dm}
\end{eqnarray}
For the $XXZ$ spin chain, the element $f=0$. In Eq. (\ref{dm}), $A,\;B$ represent the two individual qubits and $z,\;f$ are real numbers. The eigenvalues
of $\rho_{AB}$ are \cite{sarandy}
\begin{eqnarray}
 \lambda_{0}&=&\frac{1}{4}\left\{\left(1+c_{3}\right)+\sqrt{4 c_{4}^{2}+\left(c_{1}-c_{2}\right)^{2}}\right\} \nonumber \\
 \lambda_{1}&=&\frac{1}{4}\left\{\left(1+c_{3}\right)-\sqrt{4 c_{4}^{2}+\left(c_{1}-c_{2}\right)^{2}}\right\} \nonumber \\
 \lambda_{2}&=&\frac{1}{4}\left(1-c_{3}+c_{1}+c_{2}\right) \nonumber \\
 \lambda_{3}&=&\frac{1}{4}\left(1-c_{3}-c_{1}-c_{2}\right)
 \label{ev}
\end{eqnarray}
with 
\begin{eqnarray}
 c_{1}&=&2z+2f \nonumber \\
 c_{2}&=&2z-2f \nonumber \\
 c_{3}&=&a+d-2b \nonumber \\
 c_{4}&=&a-d 
 \label{cv}
\end{eqnarray}
The mutual information $I\left(\rho_{AB}\right)$ can be written as \cite{sarandy}
\begin{eqnarray}
 I\left(\rho_{AB}\right)=S\left(\rho_{A}\right)+S\left(\rho_{B}\right)
                         +\sum_{\alpha=0}^{3}\lambda_{\alpha}\log_{2}\lambda_{\alpha}
\label{mi}
\end{eqnarray}
where 
\begin{eqnarray}
 S\left(\rho_{A}\right)=S\left(\rho_{B}\right)=-\frac{1+c_{4}}{2}\log_{2}\frac{1+c_{4}}{2}
-\frac{1-c_{4}}{2}\log_{2}\frac{1-c_{4}}{2}
\label{vne}                                               
\end{eqnarray}
With expressions for $I\left(\rho_{AB}\right)$ and $C\left(\rho_{AB}\right)$ given in Eqs. 
(\ref{mi}), (\ref{vne}) and (\ref{class}), the QD, $Q\left(\rho_{AB}\right)$, (Eq. (\ref{qd}))
can in principle be computed. The difficulty lies in carrying out the maximization procedure needed
for the computation of $C\left(\rho_{AB}\right)$. It is possible to do so analytically when 
$\rho_{AB}$ is of the form given in Eq. (\ref{dm}) resulting in the following expressions for the 
QD \cite{fanchini}: 

\begin{eqnarray}
 Q\left(\rho_{AB}\right)=\mbox{min}\left\{Q_{1},Q_{2}\right\}
\label{qdf}
\end{eqnarray}
where 
\begin{eqnarray}
 Q_{1}=S\left(\rho_{B}\right)-S\left(\rho_{AB}\right)-a\log_{2}\frac{a}{a+b}
-b\log_{2}\frac{b}{a+b}
-d\log_{2}\frac{d}{d+b}-b\log_{2}\frac{b}{d+b}\nonumber \\
 \label{qdf1}
\end{eqnarray}
and
\begin{eqnarray}
 Q_{2}=S\left(\rho_{B}\right)-S\left(\rho_{AB}\right)-\Delta_{+}\log_{2}\Delta_{+}
-\Delta_{-}\log_{2}\Delta_{-}
 \label{qdf2}
\end{eqnarray}
with $\Delta_{\pm}=\frac{1}{2}\left(1\pm\Gamma \right)$ and $\Gamma^{2}=\left(a-d\right)^{2}+4\left(|z|+|f|\right)^{2}$ 

Let $\rho_{ij}$ be the reduced density matrix for two spins located at the site $i$ and $j$ respectively. $\rho_{ij}$ has the form 
shown in Eq. (\ref{dm}). The matrix elements of $\rho_{ij}$ can be expressed in terms of single-site magnetization and two-site 
spin correlation functions. In the case of the transverse-field $XY$ spin chain, the matrix elements are \cite{dutta,osborne,dillenschneider}
\begin{eqnarray}
 a&=&\frac{1}{4}+\frac{\langle\sigma^{z}\rangle}{2}+
     \frac{\langle\sigma_{i}^{z}\sigma_{j}^{z}\rangle}{4} \nonumber \\
 d&=&\frac{1}{4}-\frac{\langle\sigma^{z}\rangle}{2}+
     \frac{\langle\sigma_{i}^{z}\sigma_{j}^{z}\rangle}{4} \nonumber \\
 b&=&\frac{1}{4}\left(1-\langle\sigma_{i}^{z}\sigma_{j}^{z}\rangle\right) \nonumber \\
 z&=&\frac{1}{4}
     \left(\langle\sigma_{i}^{x}\sigma_{j}^{x}\rangle+\langle\sigma_{i}^{y}\sigma_{j}^{y}\rangle\right) \nonumber \\
 f&=&\frac{1}{4}
     \left(\langle\sigma_{i}^{x}\sigma_{j}^{x}\rangle-\langle\sigma_{i}^{y}\sigma_{j}^{y}\rangle\right) 
 \label{elements}
\end{eqnarray}
The single-site magnetization $\langle\sigma^{z}\rangle$ in the case of the transverse-field 
$XY$ model is given by \cite{dutta,dillenschneider} 
\begin{eqnarray}
 \langle\sigma^{z}\rangle=-\frac{1}{\pi}\int_{0}^{\pi}d\phi\frac{\left(1+\lambda\cos{\phi}\right)}
                           {\omega_{\phi}}
\end{eqnarray}
where 
\begin{eqnarray}
 \omega_{\phi}=\sqrt{\left(\gamma\lambda\sin{\phi}\right)^{2}+\left(1+\lambda\cos{\phi}\right)^{2}}
\end{eqnarray}
describes the energy spectrum. The spin-spin correlation functions are obtained from the determinants of 
Toeplitz matrices \cite{dillenschneider,mccoy,pfeuty} as
\begin{eqnarray}
  \left\langle\sigma_{i}^{x}\sigma_{i+r}^{x}\right\rangle &=& 
 \left|
 \begin{array}{cccc}
  G_{-1} & G_{-2} & \cdots & G_{-r}\\
  G_{0} & G_{-1} & \cdots & G_{-r+1}\\
  \vdots & \vdots & \ddots & \vdots\\
  G_{r-2} & G_{r-3} & \cdots & G_{-1}\\
 \end{array}\right| \nonumber \\
  \left\langle\sigma_{i}^{y}\sigma_{i+r}^{y}\right\rangle &=& 
 \left|
 \begin{array}{cccc}
  G_{1} & G_{0} & \cdots & G_{-r+2}\\
  G_{2} & G_{1} & \cdots & G_{-r+3}\\
  \vdots & \vdots & \ddots & \vdots\\
  G_{r} & G_{r-1} & \cdots & G_{1}\\
 \end{array}\right| \nonumber \\
 \left\langle\sigma_{i}^{z}\sigma_{i+r}^{z}\right\rangle &=&
 \left\langle\sigma^{z}\right\rangle^{2}-G_{r}G_{-r}
\label{toepliz} 
\end{eqnarray}
where 
\begin{eqnarray}
  G_{r}=\frac{1}{\pi}\int_{0}^{\pi}d\phi\cos(r\phi)
 \frac{\left(1+\lambda\cos\phi\right)}{\omega_{\phi}}
-\frac{\gamma\lambda}{\pi}\int_{0}^{\pi}d\phi\sin(r\phi)\frac{\sin\phi}{\omega_{\phi}}
 \label{spincor}
\end{eqnarray}
with $r=|i-j|$ being the distance between the two spins at the sites $i$ and $j$ (for the 
nearest-neighbour case, $r=1$). 

\begin{figure}
\begin{center}
 \includegraphics[scale=0.6]{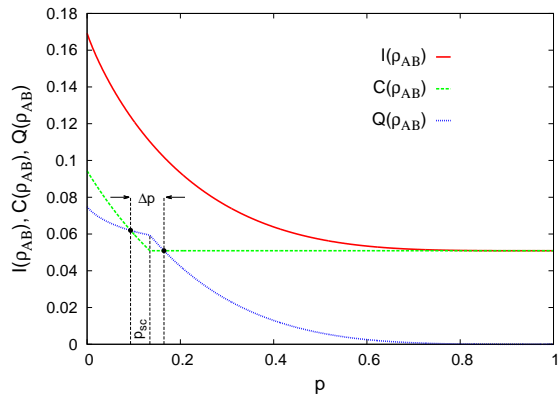}
\end{center}
\caption{PF channel: Decay of mutual information 
         $I\left(\rho_{AB}\right)$ (solid line), classical correlations $C\left(\rho_{AB}\right)$
         (dashed line), and the QD $Q\left(\rho_{AB}\right)$ (dot-dashed line) as a function of the parametrized time 
         $p=1-e^{-\theta t}$ for $\lambda=0.5$ and $\gamma=1$. $Q\left(\rho_{AB}\right)$ is greater than $C\left(\rho_{AB}\right)$ in the parametrized time interval $\Delta p$.}
\label{pfd}
\end{figure}

\section{Results}

Three general types of dynamics under the effect of decoherence have been observed in an earlier study \cite{maziero5}: \textit{(i)} $C\left(\rho_{AB}\right)$ 
is constant as a function of time and $Q\left(\rho_{AB}\right)$ decays monotonically, \textit{(ii)}
$C\left(\rho_{AB}\right)$ decays monotonically over time till a parametrized time $p_{sc}$ is reached
and remains constant thereafter. At $p_{sc}$, $Q\left(\rho_{AB}\right)$ has an abrupt change in the decay 
rate which has been demonstrated in actual experiments \cite{xu,auccaise}. 
Also, a parametrized time interval exists in which $Q\left(\rho_{AB}\right)$ has 
a magnitude greater than that of $C\left(\rho_{AB}\right)$ and \textit{(iii)} both 
$C\left(\rho_{AB}\right)$ and $Q\left(\rho_{AB}\right)$ decay monotonically. Mazzola et al. \cite{mazzola1}
have obtained the significant result that under Markovian dynamics and for a class of initial states 
the QD remains constant in a finite time interval $0<t<\tilde{t}$. In this time interval, the classical 
correlations $C\left(\rho_{AB}\right)$ decay monotonically. Beyond $t=\tilde{t}$, 
$C\left(\rho_{AB}\right)$ becomes constant whereas the QD decreases monotonically as a function of 
time. In the case of the spin models under consideration,  
the rules of evolution for the coefficients are obtained from Eqs. (\ref{dm}), (\ref{elements}) and (\ref{evolve}) with the 
choice of the Kraus operators dictated by the specific type of quantum channel (see Eqs. (\ref{kbf}), (\ref{kpf}) and (\ref{kbpf})). The evolution rules
are given by
 
\noindent\textit{BPF Channel:}
\begin{eqnarray}
 c_{1}(p)&=&c_{1}(0)(1-p)^{2} \nonumber\\
 c_{2}(p)&=&c_{2}(0)\nonumber\\
 c_{3}(p)&=&c_{3}(0)(1-p)^{2} \nonumber\\
 c_{4}(p)&=&c_{4}(0)(1-p)
\label{bpft}
\end{eqnarray}
\noindent\textit{BF Channel:}
\begin{eqnarray}
 c_{1}(p)&=&c_{1}(0)\nonumber\\
 c_{2}(p)&=&c_{2}(0)(1-p)^{2}\nonumber\\
 c_{3}(p)&=&c_{3}(0)(1-p)^{2} \nonumber\\
 c_{4}(p)&=&c_{4}(0)(1-p)
\label{bft}
\end{eqnarray} 
\noindent\textit{PF Channel:}
\begin{eqnarray}
 c_{1}(p)&=&c_{1}(0)(1-p)^{2}\nonumber\\
 c_{2}(p)&=&c_{2}(0)(1-p)^{2}\nonumber\\
 c_{3}(p)&=&c_{3}(0) \nonumber\\
 c_{4}(p)&=&c_{4}(0)
\label{pft}
\end{eqnarray}

\begin{figure}
\begin{center}
 \includegraphics[scale=0.6]{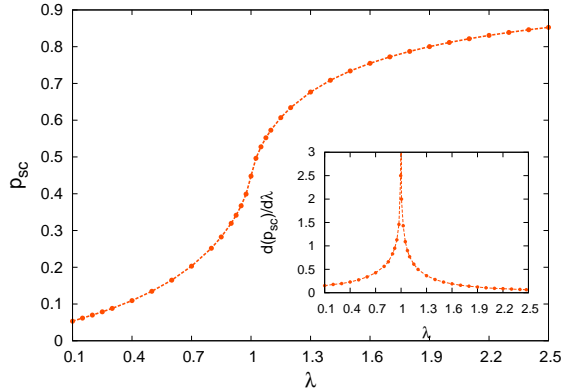}
\end{center}
\caption{Variation of $p_{sc}$ as a function of $\lambda$ with $\gamma=1$ for the PF channel. (inset) The first derivative of $p_{sc}$ w.r.t. $\lambda$ diverges as 
         the QCP $\lambda_{c}=1$ is approached.}
\label{pfl}
\end{figure}

For the transverse-field $XY$ model, the dynamical evolution of the mutual information $I\left(\rho_{AB}\right)$, the classical correlations $C\left(\rho_{AB}\right)$ and the QD $Q\left(\rho_{AB}\right)$
in the case of the BPF channel are shown as a function of the parametrized time $p\;\left(p=1-e^{-\theta t}\right)$ in figure \ref{dyn} for $\lambda=\gamma=0.7$. The dynamics are similar to type \textit{(ii)} where sudden changes in the decay rates of 
the classical correlations and the QD occur at a parametrized time instant $p_{sc}$. The inset of the figure \ref{dyn} shows the time evolution of $I\left(\rho_{AB}\right)$, $C\left(\rho_{AB}\right)$ 
and $Q\left(\rho_{AB}\right)$ for $\lambda=0.7$, $\gamma=-0.7$ which are described by the type \textit{(iii)} dynamics, i.e. both $C\left(\rho_{AB}\right)$ and $Q\left(\rho_{AB}\right)$ decay monotonically.
$Q\left(\rho_{AB}\right)$ tends to zero in the asymptotic limit $p\rightarrow 1$ whereas $C\left(\rho_{AB}\right)$ and $I\left(\rho_{AB}\right)$ have finite values in the same limit. 
The dynamics in the case of the BF channel have an interesting 
correspondence with the dynamics of the BPF channel. Type \textit{(ii)} and Type \textit{(iii)} 
dynamics are obtained in the case of the BPF (BF) channel for +ve (-ve) and -ve (+ve) values 
of $\gamma$ respectively \cite{pal3}. An analytical expression of the parametrized time $p_{sc}$ at which the sudden changes in the decay rate of  $C\left(\rho_{AB}\right)$ and $Q\left(\rho_{AB}\right)$ take place
can be obtained for the BPF/BF channels from Eqs. (\ref{bpft}) and (\ref{bft}) respectively \cite{pal3}. The dynamics of $I\left(\rho_{AB}\right)$, $C\left(\rho_{AB}\right)$ and 
$Q\left(\rho_{AB}\right)$ in the case of the PF channel belong to type \textit{(ii)} with $Q\left(\rho_{AB}\right)>C\left(\rho_{AB}\right)$ during a parametrized time interval \cite{pal3}.  
The classical correlations remain 
constant at a value $I\left(\rho_{AB}\right)_{p=1}$, the mutual information of the fully decohered state,
in the parametrized time interval $p_{sc}< p<1$. In the interval $0<p< p_{sc}$, the 
classical correlations decay with time. On the other hand, the quantum correlation, as measured by 
the QD, undergoes a sudden change in its decay rate at $p=p_{sc}$ and goes to zero 
in the asymptotic limit $p\rightarrow1$. The type of dynamics in the case of the 
PF channel remains the same under a change in sign of the anisotropy parameter $\gamma$.

Figure \ref{qpt} shows the plots of $p_{sc}$ and the first derivative of $p_{sc}$ w.r.t. $\lambda$ (inset) as functions of $\lambda$
for the BPF channel $(\gamma>0)$. The first derivative of $p_{sc}$ w.r.t. $\lambda$ diverges as the QCP $\lambda_{c}=1$ is approached indicating a QPT.
In the case of the PF channel, the dynamics are of type \textit{(ii)} and $Q\left(\rho_{AB}\right)$ is greater than $C\left(\rho_{AB}\right)$ in a parametrized time interval. 
Also, the first derivative of $p_{sc}$ w.r.t. $\lambda$ for a fixed value of the anisotropy parameter $\gamma$ (both $+$ve and $-$ve) diverges at the QPT point $\lambda_{c}=1$ \cite{pal3}. 
On the other hand,  the first derivative of $p_{sc}$ w.r.t. $\gamma$ for a fixed value of $\lambda$ 
shows a discontinuity at the anisotropy transition point $\gamma=0$ indicating a QPT for all the three channels BPF, BF and PF \cite{pal3}. In the 
case of the PF channel, $p_{sc}$ has a symmetric variation  across the anisotropy transition point $\gamma=0$ whereas a parametric time $p_{sc}$ exists only for $\gamma>0$ ($\gamma<0$)
in the case of the BPF (BF) channel. A fuller description and discussion of the results on the dynamics of the classical and quantum correlations are given in Ref. [\refcite{pal3}].
In the case of the AD channel, the dynamics are of type \textit{(iii)} with all the three correlations, $I\left(\rho_{AB}\right)$, $C\left(\rho_{AB}\right)$ and $Q\left(\rho_{AB}\right)$ decaying 
asymptotically $(p\rightarrow 1)$ to zero. In the case of the TIM in 1d, 
the signature of quantum criticality at $\lambda_{c}=1$ is obtained only in the cases of the BPF and PF channels. In the case of the PF channel, the dynamics are of type 
\textit{(ii)} (figure \ref{pfd}). Figure \ref{pfl} shows the variation of $p_{sc}$ as a function of the tuning parameter $\lambda$. The inset of the figure shows that the first derivative of $p_{sc}$ w.r.t. 
$\lambda$ exhibits a divergence as the quantum critical point $\lambda_{c}=1$ is approached.

\begin{figure}
\begin{center}
 \includegraphics[scale=0.3]{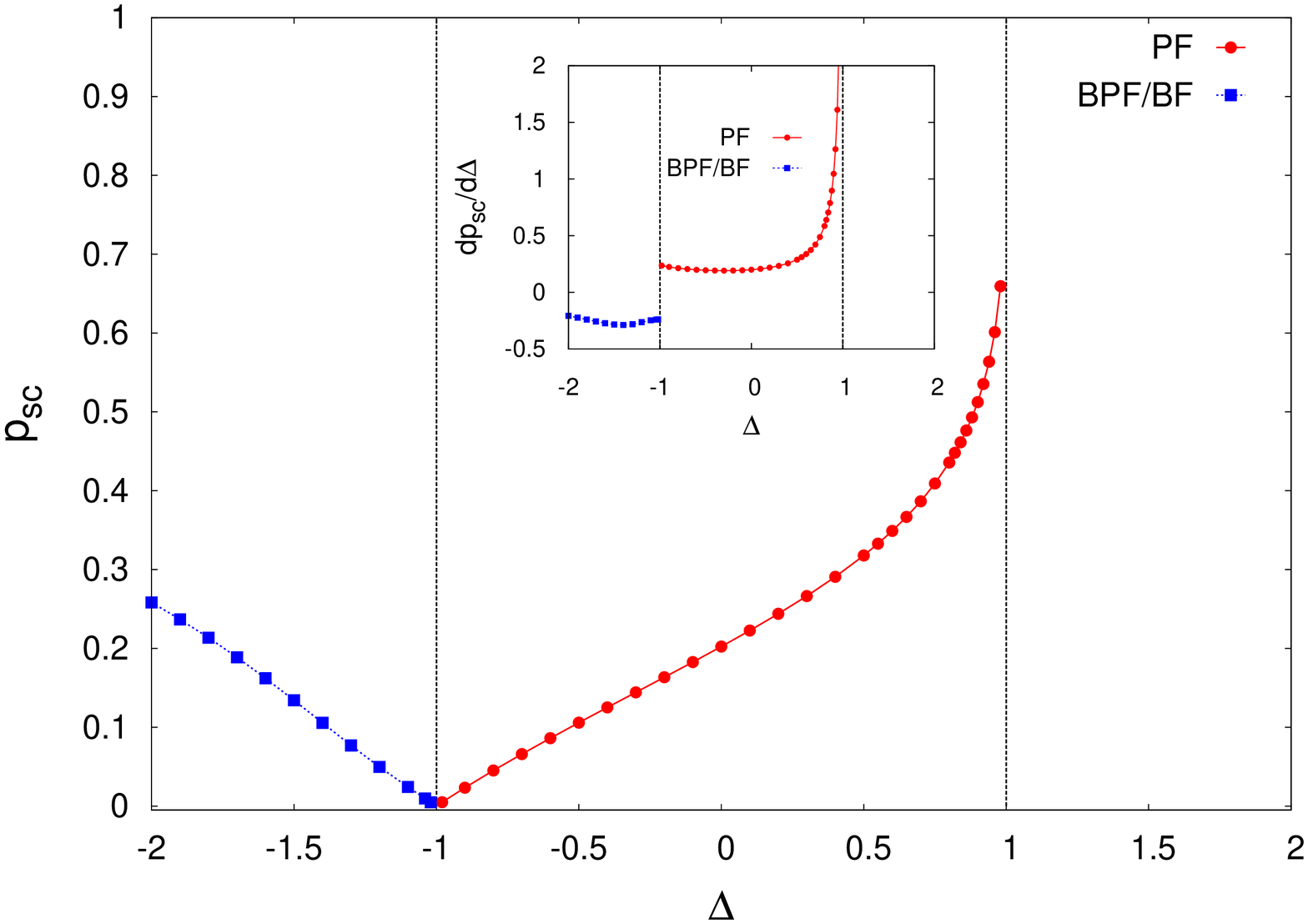}
\end{center}
\caption{Variation of $p_{sc}$ as a function of $\Delta$ for the BPF/BF ($\Delta<-1$) (line with solid squares) and the PF ($-1<\Delta<1$) (line with solid circles) channel. 
(inset) In the case of the PF channel, the first derivative of $p_{sc}$ w.r.t. $\Delta$ diverges as the ferromagnetic point $\Delta=1$ is approached.}
\label{xxz}
\end{figure}

In the case of the $XXZ$ spin chain, the critical point analysis in the presence of decoherence is carried out utilizing the results obtained in Refs. [\refcite{sarandy,maziero5}]. 
Figure \ref{xxz} exhibits 
the variation of $p_{sc}$ as a function of the anisotropy constant  $\Delta$ (Eq. (\ref{hxxz2})) in the cases of the BPF/BF (solid squares) and the PF (solid circles) channels. 
In the case of the PF channel, the first derivative $\frac{dp_{sc}}{d\Delta}$ diverges as the ferromagnetic 
point $\Delta=1$ is approached. One also notes that $p_{sc}$ has a finite value when $\Delta$ is $<-1$ ($-1<\Delta<1$) in the case of the BF/BPF (PF) channel. 

\section{Concluding Remarks} 
In this review, we have considered three spin models, namely, the TIM in 1d, the transverse-field $XY$ and the $XXZ$ spin chains. 
The two-spin reduced density matrix obtained from the 
ground state in each case has the general form of $X$-states (further simplified in the case of the $XXZ$ chain). 
The matrix elements of the reduced density matrix can be expressed in terms of 
single-spin expectation values and two-spin correlation functions which are mostly known. Analytical 
formulae are available to compute the mutual information $I\left(\rho_{AB}\right)$, the classical correlations $C\left(\rho_{AB}\right)$ 
and the QD $Q\left(\rho_{AB}\right)$ in the two-spin state described by the reduced density matrix \cite{sarandy,luo,adesso,ali}. 
In the presence of system-environment interactions resulting in decoherence, the time evolution of the reduced density matrix can be determined 
in the Kraus operator formalism \cite{nielsen,auccaise}. The spin models under consideration exhibit QPTs at specific values of a tuning parameter, e.g., 
magnetic field strength or an anisotropy parameter. While there is a large number of studies on how 
quantum correlation measures provide signatures of QPTs \cite{vedralRMP,senAP,modi3,celeri,bose}, there is little investigation 
so far on appropriate indicators of QPTs in the presence of decoherence
\cite{pal2,pal3}. In the case of Markovian time evolution, quantities like $\frac{dp_{sc}}{d\lambda}$ and $\frac{dp_{sc}}{d\gamma}$
(transverse-field $XY$ model), associated with type \textit{(ii)}
dynamics, diverge/become discontinuous as a quantum critical point is approached. Dynamics similar to type \textit{(ii)} are associated with specific quantum channels, e.g., in 
the case of the AD channel the dynamics are of type \textit{(iii)} which do not provide any indication of the occurrence of a QPT. There are several directions in which the 
investigations reported in Refs. [\refcite{pal2,pal3}] can be extended. The analysis has so far been carried out for the $X$-states but since an analytic computational scheme 
for a general two-qubit state is now available \cite{adesso}, one 
could probe the dynamics of classical and quantum correlations in general two-spin states and look for signatures of QPTs. 
In Ref. [\refcite{pal2}], the reduced 
density matrix describes n.n. spin pairs and the analysis is carried out at $T=0$. Extension to further-neighbours and 
finite temperature cases has been carried out in Ref. [\refcite{pal3}]. 
The more general case of non-Markovian time evolution 
vis-\`{a}-vis signatures of QPTs is yet to be addressed. The dynamics of multipartite quantum correlations constitute another important area of study. Models of 
interacting fermions and 
bosons exhibit a variety of QPTs \cite{vedralRMP,senAP} which could be signaled by quantities associated with the dynamics of quantum correlations. Finally, 
recent experiments 
\cite{xu,auccaise} on the dynamics of quantum correlations set new challenges in exploring the interconnections between quantum correlations, decoherence and 
quantum phase transitions.


\section*{References}
\begin{thebibliography}{100}
\bibitem{epr}A. Einstein, B. Podolsky, and N. Rosen, \textit{Phys. Rev.} \textbf{47}, 777 (1935)
\bibitem{vedralRMP} L. Amico, R. Fazio, A. Osterloh and V. Vedral \textit{Rev. Mod. Phys. } \textbf{80}, 517 (2008)
\bibitem{senAP}M. Lewenstein, A. Sanpera, V. Ahufinger, B. Damski, A. Sen and U. Sen, \textit{Adv. in Phys.} \textbf{56}, 243 (2007) 
\bibitem{latorre}J. I. Latorre and A. Riera \textit{J. Phys.} \textbf{A42}, 504002 (2009) 
\bibitem{olivier} H. Olivier and W. H. Zurek \textit{Phys. Rev. Lett.} \textbf{88}, 017901 (2001)
\bibitem{zurekRMP} W. H. Zurek, \textit{Rev. Mod. Phys.} \textbf{75}, 715 (2003) 
\bibitem{henderson} L. Henderson and V. Vedral, \textit{J. Phys.} \textbf{A34} 6899 (2001)
\bibitem{nielsen}M. A. Nielsen and I. L. Chuang, \textit{Quantum Computation and Quantum Information} (Cambridge University Press, Cambridge, 2000)
\bibitem{sarandy} M. S. Sarandy, \textit{Phys. Rev. } \textbf{A80} 022108 (2009)
\bibitem{luo}S. Luo, \textit{Phys. Rev.} \textbf{A77} 042303 (2008) 
\bibitem{adesso}D. Girolami and G. Adesso, \textit{Phys. Rev. } \textbf{A83}, 052108 (2011)
\bibitem{ali} M. Ali, A. R. P. Rau and G. Alber, \textit{Phys. Rev.} \textbf{A81}, 042105 (2010)
\bibitem{modi}K. Modi, T. Paterek, W. Son, V. Vedral and M. Williamson, \textit{Phys. Rev. Lett.} \textbf{104}, 080501 (2010)
\bibitem{modi2}K. Modi and V. Vedral, AIP Conf. Proc. \textbf{1384}, 69-75 (2011)
\bibitem{dakic}B. Daki\'{c}, V. Vedral and C. Brukner, \textit{Phys. Rev. Lett.} \textbf{105}, 190502 (2010)
\bibitem{giorda} P. Giorda  and M. G. A. Paris, \textit{Phys. Rev. Lett.} \textbf{105}, 020503 (2010)
\bibitem{modi3}K. Modi, A. Brodutch, H. Cabb, T. Paterek and V. Vedral, quant-ph/1112.6238v1 (2011)

\bibitem{celeri} L. C. C\'{e}leri, J. Maziero and R. M. Serra, \textit{Quantum Information} \textbf{9}, 1837 (2011)
\bibitem{dillenschneider}R. Dillenschneider, \textit{Phys. Rev. }\textbf{A78}, 224413 (2008)
\bibitem {pal1}A. K. Pal and I. Bose, \textit{J. Phys. }\textbf{B44}, 045101 (2011)
\bibitem{werlang}T. Werlang and G. Rigolin, \textit{Phys. Rev. }\textbf{A81}, 044101 (2010)
\bibitem{maziero2}J. Maziero, H. C. Guzman, L. C. C\'{e}leri,  M. S. Sarandy and R. M. Serra, \textit{Phys. Rev. }\textbf{A82}, 012106 (2010)
\bibitem{werlang2} T. Werlang, C. Trippe, G. A. P. Ribeiro and G. Rigolin, \textit{Phys. Rev. Lett.} \textbf{105}, 095702 (2010)

\bibitem{maziero3} J. Maziero, L. C. C\'{e}leri, R. M. Serra and M. S. Sarandy, \textit{Phys. Lett. } \textbf{A376}, 1540 (2012)
\bibitem{libherti} L. Ciliberti, R. Rosignoli and N. Canosa, \textit{Phys. Rev. } \textbf{A82} 042316 (2010)
\bibitem{hassan} A. S. M. Hassan, B. Lari and P. S. Joag, \textit{J. Phys} \textbf{A43} 485302 (2010)

\bibitem{tomasello} B. Tomasello, D. Rossini, A. Hamma and L. Amico, \textit{Europhys. Lett.} \textit{96}, 27002  (2011)

\bibitem{dhar}H. S. Dhar, R. Ghosh, A. Sen (De) and U. Sen, \textit{Europhys. Lett.} \textit{98}, 30013 (2012)
\bibitem{pal2}A. K. Pal and I. Bose, \textit{Eur. Phys. J.} \textbf{B85}: 36 (2012)
\bibitem{chen}Y. -X. Chen and Z. Yin, \textit{Comm. Theor. Phys.} \textbf{54}, 60 (2010)
\bibitem{tian} L. -J. Tian, Y. -Y. Yan and L.-G. Qin, quant-ph/1104.1525v2
\bibitem{yurischev}M. A. Yurischev, \textit{Phys. Rev. } \textbf{B84}, 024418 (2011)
\bibitem{sachdev}S. Sachdev, \textit{Science} \textbf{288}, 475 (2000)
\bibitem{sachdev2}S. Sachdev, \textit{Quantum Phase Transitions} (Cambridge University Press, Cambridge, 1999)
\bibitem{maziero4} J. Maziero, T. Werlang, F. F. Fanchini, L. C. C\'{e}leri, R. M. Serra, \textit{Phys. Rev.} \textbf{A81}, 022116 (2010)
\bibitem{almeida} M. P. Almeida et al., \textit{Science} \textbf{316}, 579 (2007)
\bibitem{werlang3} T. Werlang, S. Souza, F. F. Fanchini, C. Villas Boas, \textit{Phys. Rev.} \textbf{A80}, 024103 (2009)
\bibitem{maziero5} J. Maziero, L. C. C\'{e}leri, R. M. Serra, V. Vedral, \textit{Phys. Rev.} \textbf{A80}, 044102 (2009)
\bibitem{mazzola1} L. Mazzola, J. Piilo, S. Maniscalco, \textit{Phys. Rev. Lett.} \textbf{104}, 200401 (2010)

\bibitem{pal3}A. K. Pal and I. Bose, \textit{Eur. Phys. J.} \textbf{B85}:277 (2012)
\bibitem{ferraro}A. Ferraro, L. Aolita, D. Cavalcanti, F. M. Cacchietti and A. Acin, \textit{Phys. Rev.} \textbf{A81}, 052318 (2010)
\bibitem{mattis} E. Lieb, T. Schultz and D. Mattis, \textit{Ann. Phys.} \textbf{60}, 407 (1961)
\bibitem{mccoy} E. Barouch and B. McCoy, \textit{Phys. Rev.} \textbf{A3}, 786 (1971)
\bibitem{pfeuty} P. Pfeuty, \textit{Ann. Phys.} \textbf{57}, 79 (1970)
\bibitem{zhong} M. Zhong and P. Tong, \textit{J. Phys.} \textbf{A43}, 505302 (2010)

\bibitem{dutta}A. Dutta, U. Divakaran, B. K. Chakrabarti, T. F. Rosenbaum and G. Aeppli, cond-mat.stat-mech/1012.0653v1 
\bibitem{its}A. R. Its, B.-Q. Jin, V. E. Korepin, Journal Phys. \textbf{A38}, 2975-2990, (2005)
\bibitem{osborne} T. Osborne and M. A. Nielsen, \textit{Phys. Rev.} \textbf{A66}, 032110 (2002)
\bibitem{osterloh}A. Osterloh, L. Amico, G. Falci and R. Fazio, \textit{Nature} \textbf{416}, 608 (2002)
\bibitem{bose}I. Bose and A. Tribedi in \textit{Quantum Quenching, Annealing and Computation} ed. A. K. Chandra, A. Das and B. K. Chakrabarti (Springer, Heidelberg, 2010) p. 177, Chapter 8
\bibitem{zanardi}P. Zanardi and N. Paunkovi\'{c}, \textit{Phys. Rev. } \text{E74}, 031123 (2006)
\bibitem{tribedi} A. Tribedi and I. Bose, \textit{Phys. Rev. } \textbf{A79}, 012331 (2009)
\bibitem{salles}A. Salles et. al., \textit{Phys. Rev.} \textbf{A78}, 022322 (2008)
\bibitem{fanchini} F. F. Fanchini, T. Werlang, C. A. Brasil, L. G. E. Arruda and A. O. Caldeira, \textit{Phys. Rev.} \textbf{A81}, 052107 (2010) 
\bibitem{xu}J. -S. Xu, X. -Y. Xu, C. -F. Li, C. -J. Zhong, X. -B. Zoa and G. -C. Guo, \textit{Nat. Commun.} \textbf{1}, 7 (2010)
\bibitem{auccaise} R. Auccaise et al., \textit{Phys. Rev. Lett.} \textbf{107}, 140403 (2011) 
\end{thebibliography}
\end{document}